\documentclass[11pt,a4paper]{amsart}
\usepackage{graphics}

\newtheorem{Def}{Definition}
\newtheorem{Rem}[Def]{Remark}
\newtheorem{Lem}[Def]{Lemma}
\newtheorem{The}[Def]{Theorem}

\newtheorem{Pro}[Def]{Proposition}
\newtheorem{Exa}[Def]{Example}

\begin{document}
\title[Star-free sets of integers]
{Towards a characterization of the star-free sets of integers}
\author[M. Rigo]{Michel Rigo}
\address{Michel Rigo\newline
Institut de Math\'ematique\newline
Universit\'e de Li\`ege\newline
Grande Traverse 12 (B 37)\newline
B-4000 Li\`ege\newline
Belgium}
\email{M.Rigo@ulg.ac.be}
\date{\today}

\begin{abstract} 
    Let $U$ be a numeration system, a set $X\subseteq \mathbb{N}$ is 
    $U$-star-free if the 
    set made up of the $U$-representations of the elements in $X$ is 
    a star-free regular language. Answering a question of A.~de Luca and 
    A.~Restivo \cite{dLR}, 
    we obtain a complete logical characterization of the $U$-star-free sets of 
    integers for suitable numeration systems related to a Pisot 
    number and in particular for integer base systems. For these 
    latter systems, we study as well the problem of the base dependence. 
    Finally, the case of $k$-adic systems is also investigated.
\end{abstract}
\maketitle

\section{Introduction} 
In the study of numeration systems, a natural question is to 
determine if a set of non-negative integers has a ``simple" representation within the 
considered number system. Otherwise stated, is it possible for 
a given set $X\subset\mathbb{N}$, to find a 
``simple'' algorithm (a finite automaton) testing membership in $X$ ? 
This question has given rise to a lot of papers 
dealing with the so-called 
{\it recognizable} sets. A subset $X$ of $\mathbb{N}$ is 
said to be  $k$-recognizable if the language made up of the 
$k$-ary expansions of all the elements in $X$ is regular 
(i.e., recognizable by a finite automaton). 

Since the work of A.~Cobham \cite{Cob}, it is well-known that the recognizability 
of a set depends on the base of the numeration system. If $k$ and $l$ are two 
multiplicatively independent integers then the only subsets of 
$\mathbb{N}$ which are simultaneously 
$k$-recognizable and $l$-recognizable are exactly the ultimately 
periodic sets.

Among the recognizable sets, it could be interesting to describe the 
sets whose corresponding languages of representations belong 
to a specific subset of regular languages. Among 
the regular languages, the ``simplest" are certainly the 
star-free languages because the automata accepting those languages 
are counter-free. 
Having in mind this idea of ``simpler" representation of a set, A.~de 
Luca and A.~Restivo have considered in 
\cite{dLR} the problem of 
determining the existence of a suitable base $k$ such that the $k$-ary 
representations of the elements belonging to a set $X\subset \mathbb{N}$ 
made up a regular language of (unrestricted) 
star-height $0$ (such a set is then said to be {\it $k$-star-free}). 
One of the main results of 
\cite{dLR} is that if a $l$-recognizable set $X$ is such that its density 
function is bounded by $c (\log n)^d$, for some  
constants $c$ and $d$, then there exists a base $k$ such that $X$ is 
$k$-star-free.  

The star-free languages have been extensively studied in the 
literature \cite{NP,S,STT,T}. 
In particular M.P.~Sch\"utzenberger has shown that the star-free languages 
--- i.e., the languages expressed in terms of extended regular 
expressions without the star operation --- are exactly the aperiodic 
languages \cite{S}. 
We recall that a language $L\subset \Sigma^*$ is 
 {\it aperiodic} if there 
exists a positive integer $n$ such that for all words $u,v,w \in 
\Sigma^*$, $$uv^nw \in L \Leftrightarrow uv^{n+1}w \in L.$$

In the present paper, we answer some of the remaining open questions 
adressed in \cite{dLR} about 
sets of integers having a representation of star height $0$. 
Especially, we give a complete characterization of the $k$-recognizable sets  
 such that the language of $k$-ary expansions is aperiodic. To obtain 
 this result, we use the first-order logical characterization of the star-free languages 
 given by R. McNaughton and S. Papert \cite{NP}. 
 
 In the first two sections, 
 for the sake of simplicity  
 we consider the case of the binary system. Next, we show how our  
 results can be extended not only to the $k$-ary systems but also to numeration systems 
 defined by a linear recurrent sequence whose characteristic 
 polynomial is the minimal polynomial of a Pisot number (a Pisot 
 number is an algebraic integer $\theta>1$ such that the other roots 
 of the minimal polynomial of $\theta$ have modulus less than one). In this wider 
 framework, we have to consider the additional assumption that the set 
 of all the representations computed by the greedy algorithm is 
 star-free. For instance, this assumption is satisfied for the 
 Fibonacci system. In Section \ref{sec5}, we consider the problem of 
 the base dependence of the aperiodicity of the representations 
 for integer base systems. The obtained result 
 can be related to the celebrated Cobham's theorem: only ultimately 
 periodic sets can be $k$- and $l$-star-free for two 
 multilplicatively integers $k$ and $l$ but if the period $p$ of an 
 ultimately periodic set is 
 greater than $1$ then this set is $k$-star-free for some $k$ 
 depending on $p$ but not for all $k\in\mathbb{N}$. In particular, we 
 show that a set is $k$-star-free if and only if it is 
 $k^n$-star-free for any $n\ge 1$. In the last section, we consider 
 the case of the unambiguous $k$-adic numeration system. It is worth noticing 
 that the unique $k$-adic 
 representation of an integer is not computed through the greedy 
 algorithm and therefore this system differs from the other systems 
 encountered in this paper. It appears unsurprisingly 
 that the star-free sets with respect to this latter system are 
 exactly the $k$-star-free sets. 

 In the following, we assume the reader familiar with basic formal languages theory (see for 
 instance, \cite{Ei}). Finite automata will be denoted as 
 $5$-tuples $\mathcal{M}=(Q,q_0,F,\Sigma,\delta)$ where $Q$ is the set 
 of states, $q_0$ is the initial states, $F\subseteq Q$ is the set of final 
 states, $\Sigma$ is the input alphabet and $\delta:Q\times \Sigma\to 
 Q$ is the transition function.
 
\section{Logical characterization of star-free languages}
Let us consider the alphabet $\Sigma_2=\{0,1\}$. A word $w$ in $\Sigma_2^+$ can be 
identified as a finite model $\mathfrak{M}_w=(M,<,P_1)$ where $M=\{1,\ldots ,\vert w 
\vert\}$ ($\vert w \vert$ is the length of $w$), $<$ is the usual 
binary relation on $M$ and $P_1$ is a unary predicate for the set of positions 
in $w$ carrying the letter $1$. For our convenience, positions are 
counted from right to left. As an example, the word $w=1101001$ 
corresponds to the model $(M,<,P_1)$ where $M=\{1,\ldots,7\}$ and 
$P_1=\{1,4,6,7\}$. For further purposes, as in \cite{T} 
we expand this model with its 
maximal element $max$ (in the latter example, $max=7$) --- notice 
that $max$ is definable in terms of $<$. So to each 
nonempty word in $\Sigma_2^+$ is associated a model $(M,<,P_1,max)$.

A language is said to be {\it star-free} if it is obtained from finite sets 
by a finite number of Boolean operations (union, intersection and 
complementation) and concatenation products. McNaughton and Papert 
have shown that these languages are exactly those defined by 
first-order sentences when words are considered as finite ordered 
models \cite{NP} (a {\it sentence} is a formula whose all variables are 
bound). As an example, the language $1^+0^*$ is star-free because 
 if we denote  by $\overline{X}$ the complement $\Sigma_2^*\setminus X$ of 
 $X$ then 
$$ 1^+0^* =\{1\} \overline{\overline{\emptyset} \{0\} 
\overline{\emptyset}}\,   
\overline{\overline{\emptyset} \{1\} \overline{\emptyset}} \mbox{ 
where } \emptyset=\{0\}\cap \{00\}$$
and this language is also defined by the formula 
\begin{equation}\label{ex1}
    (\exists x) [P_1(x) \wedge (\forall y) (x<y \to P_1(y)) \wedge 
(\forall y)(y<x \to \neg P_1(y))].
\end{equation}
The language of all the formulas defining star-free languages will be 
denoted by $\mathcal{L}_{SF}$ (if necessary, to recall the alphabet 
$\Sigma_2$ we can write $\mathcal{L}_{SF,2}$). Notice that with these finite models, 
we are not considering the empty word.

To be precise, if $\varphi(x_0,\ldots,x_n)$ is a formula having at 
most $x_0,\ldots,x_n$ as free variables, the interpretation of 
$\varphi$ in a word-model $\mathfrak{M}_w$ having $M$ as domain and 
$r_0,\ldots ,r_n$ as $M$-elements is defined in a natural manner and 
we write $\mathfrak{M}_w\models \varphi[r_0,\ldots ,r_n]$ if $\varphi$ 
is satisfied in $\mathfrak{M}_w$ when interpreting $x_i$ by $r_i$. The 
language defined by a formula $\varphi$ is
$$\{w\in\Sigma_2^+\mid \mathfrak{M}_w\models \varphi\}.$$
\subsection{Syntax of logical formulas in 
$\mathcal{L}_{SF}$}\label{sec:SF} 
The first-order language of the finite ordered models representing 
words is defined as follows. The {\it variables} are denoted 
$x,y,z,\ldots$ and are ranging over $M$-elements. The {\it terms} are 
obtained from the variables and the constant $max$. 
The {\it atomic formulas} are obtained by the following rules: 
\begin{itemize}
\item[1.] 
if $\tau_1$ and $\tau_2$ are terms then $\tau_1 < \tau_2$ and 
$\tau_1=\tau_2$ are atomic formulas
\item[2.] if $\tau$ is a term then $P_1(\tau)$ is an atomic formula.
\end{itemize}
Finally, we obtain the set $\mathcal{L}_{SF}$ 
of all the {\it formulas} by using the 
Boolean connectives $\neg$, $\wedge$, $\vee$, $\to$, $\leftrightarrow$ 
and the first-order quantifiers $(\exists x)\ldots$ and $(\forall 
x)\ldots$ where $x$ is a variable.

Notice that for convenience we are somehow redundant in our definitions, 
$\varphi \vee \psi$ stands for $\neg (\neg \varphi \wedge \neg \psi)$, 
$x=y$ stands for $\neg((x<y)\vee (y<x))$, 
$\varphi \to \psi$ stands for $\neg \varphi \vee \psi$ and 
$\varphi \leftrightarrow \psi$ stands for $(\varphi \to \psi) \wedge 
(\psi \to \varphi)$. We also write $x\le y$ for $(x<y) \vee (x=y)$.

It is worth noting that in this formalism we can define $y=x+1$, where 
$x$ and $y$ are variables,
$$y=x+1 \equiv (x<y) \wedge (\forall z) (x<z \to y\le z)$$
but the form $z=x+y$ is not allowed, if $x$, $y$ and $z$ are variables 
\cite{NP}.

\section{Logical characterization of recognizable sets of integers}
In the present section, we consider the binary numeration system. If 
$x$ is an non-negative integer, the binary expansion of $x$ computed 
through the greedy algorithm (the {\it normalized $2$-representation} of 
$x$) is denoted $\rho_2(x)$ (for a presentation of the greedy 
algorithm, see \cite{F}). Notice that 
$\rho_2(0)$ is the empty word $\varepsilon$ and we allow leading zeroes 
in normalized $2$-representations. Thus, the set of all the normalized 
representations is
$$\mathcal{N}_2=0^*\{\rho_2(n)\mid n\in\mathbb{N}\}.$$

A set $X$ of integers is said to be {\it $2$-recognizable} if the set 
$\rho_2(X)$ of the normalized $2$-representations of all the elements in 
$X$ is a regular language.  

\begin{Rem}\label{zero}{\rm Allowing leading zeroes does not change the star-free 
behavior of a language made up of representations. Indeed, let $\Sigma_k=\{0,\ldots,k-1\}$, 
$k\ge 2$  and 
$L\subset \Sigma_k^*\setminus 0 \Sigma_k^*$ be a regular language 
consisting of 
words which do not begin with $0$. Then $L$ is star-free if and only 
if $0^*L$ is star-free. Indeed, 
$$0^*L=\overline{ 
\overline{\emptyset}\{1,\ldots,k-1\} \overline{\emptyset}} L 
\ {\rm and}\ L=0^*L\setminus (0\, \overline{\emptyset}).$$
}\end{Rem}
\begin{Def}{\rm 
A set $X\subset \mathbb{N}$ is said to be {\it 
$2$-star-free} if $\rho_2(X)$ (or equivalently $0^* \rho_2(X)$) is a regular aperiodic language.
}\end{Def}
It is well-known that the $2$-recognizable 
sets are exactly those definable in the first-order structure 
$\langle {\mathbb N},+,V_2\rangle$ (see \cite[Theorem 6.1]{BHMV} or 
\cite{Hod,V}) where $V_2(x)$ is the greatest power of $2$ dividing 
$x$ (and $V_2(0)$ is $1$). Thus $X\subset \mathbb{N}$ is said to be 
{\it $2$-definable} if there exists a formula $\varphi$ of 
$\langle {\mathbb N},+,V_2\rangle$ such that 
$$X=\{n\in \mathbb{N} 
\mid \langle {\mathbb N},+,V_2\rangle \models \varphi(n)\}.$$

Instead of $V_2(x)$, we shall use the binary relation 
$\epsilon_2(x,y)$ defined by ``$y$ is a power of $2$ occurring in the 
normalized $2$-representation of $x$''. As an example $(74,8)$ belongs 
to $\epsilon_2$ because $\rho_2(74)=100\underline{1}010$ but $(74,16)$ and 
$(74,31)$ do not. Thus we can write
$$x=\sum_{\epsilon_2(x,y)} y.$$
Observe also that $(x,x)$ belongs to $\epsilon_2$ if and only if $x$ is a 
power of $2$. 

\begin{Rem}{\rm 
The structures $\langle {\mathbb 
N},+,V_2\rangle$ and $\langle {\mathbb N},+,\epsilon_2\rangle$ are 
equivalent (i.e., for any formula $\varphi(n)$ of $\langle {\mathbb 
N},+,V_2\rangle$ there exists a formula $\varphi'(n)$ of $\langle {\mathbb 
N},+,\epsilon_2\rangle$ such that $\{n\in\mathbb{N}\mid \langle {\mathbb 
N},+,V_2\rangle \models \varphi(n)\}= \{n\in\mathbb{N}\mid \langle {\mathbb 
N},+,\epsilon_2\rangle \models \varphi'(n)\}$ and conversely). 
Indeed, $\epsilon_2(x,y)$ is defined in $\langle {\mathbb 
N},+,V_2\rangle$ by
$$ (V_2(y)=y)\wedge (\exists z)(\exists t)(x=t+y+z \wedge z<y \wedge 
(y<V_2(t) \vee t=0))$$
and $V_2(x)=y$ is defined in $\langle {\mathbb 
N},+,\epsilon_2\rangle$ by 
$$ \epsilon_2(x,y)\wedge (\forall 
z)(\epsilon_2(x,z)\to y \le z).$$
To be complete, 
notice that the binary relation $<$ is definable in  the Presburger 
arithmetic $\langle {\mathbb N},+\rangle $ by 
$$x<y \equiv (\exists z)(\neg(z=0) \wedge y=x+z).$$
}\end{Rem}

For our purposes, we introduce a subset $\mathcal{L}_{2,n}$ of 
formulas $\varphi(n)$ in $\langle {\mathbb N},+,\epsilon_2\rangle$ 
defined as follows. 
\subsection{Syntax of logical formulas in $\mathcal{L}_{2,n}$}
The {\it variables} are ranging over $\mathbb{N}$ 
and denoted $b,n,x,y,z,\ldots$ (when specified, $b$ and $n$ have some 
special role). Roughly speaking $n$ is dedicated to be the only free 
variable and $b$ plays the role of an upper limit for all the bound 
variables occurring in a formula.  
The only {\it terms} are the variables. The {\it atomic 
formulas} are obtained with the following rules:
\begin{itemize}
    \item[1.] If $x$ and $y$ are variables ($\neq b,n$) then $x<y$ and $x=y$ are 
    atomic formulas.
    \item[2.] if $x$ is a variable ($\neq b,n$) then $\epsilon_2(n,x)$ is an atomic 
    formula.
\end{itemize}
If $\varphi$ is a formula whose $x$ is a free variable ($x\neq b,n$) 
then 
$$(\exists x)_2^{<b} \varphi \equiv 
(\exists x)(\epsilon_2(x,x)\wedge x<b \wedge \varphi)$$
and 
$$(\forall x)_2^{<b} \varphi \equiv 
(\forall x)(\epsilon_2(x,x)\wedge x<b \wedge \varphi)$$
are {\it formulas}. To obtain formulas, we can also use the usual 
Boolean connectives $\neg$, $\wedge$, $\vee$, $\to$, $\leftrightarrow$ 
either for formulas or atomic formulas. 
We are now able to define $\mathcal{L}_{2,n}$. 
If $\varphi$ is a formula in which the only free variables are 
(possibly) $n$ and $b$ then 
$$(\exists b)(\epsilon_2(b,b)\wedge \varphi)$$
is a formula of $\mathcal{L}_{2,n}$ having (possibly) a single free 
variable $n$. 

\begin{Exa}{\rm 
The formula $\varphi(n)$ given by
\begin{equation}\label{ex2}
\begin{aligned}
\varphi(n) \equiv & (\exists b)\{ \epsilon_2(b,b) \wedge 
 (\exists x)_2^{<b} [\epsilon_2(n,x) 
\wedge \\ 
& (\forall y)_2^{<b} (x<y \to \epsilon_2(n,y))  \wedge 
 (\forall y)_2^{<b} (y<x \to \neg \epsilon_2(n,y)) ]\}
\end{aligned}   
\end{equation}
belongs to $\mathcal{L}_{2,n}$. We shall see that the set $X=\{n\mid \langle 
{\mathbb N},+,\epsilon_2\rangle \models \varphi(n)\}$ is such that 
$\rho_2(X)=1^+0^*$. Thus $\varphi(n)$ actually defines a $2$-star-free set of 
integers. As another example, the set $Y$ of the powers of 
$2$ is $2$-star-free because $\rho_2(Y)=10^*$ and it is also definable 
in $\mathcal{L}_{2,n}$ by the 
formula
\begin{equation}\label{ex3}
\psi(n)\equiv (\exists b)[ \epsilon_2(b,b) \wedge (\exists 
x)_2^{<b} (\epsilon_2(n,x) \wedge (\forall y)_2^{<b} 
(\epsilon_2(n,y)\to x=y))].
\end{equation}
}\end{Exa}

With this definition of $\mathcal{L}_{2,n}$, we obtain quite easily 
the following result.
\begin{The}\label{thm1} A set $X\subseteq\mathbb{N}$ is $2$-star-free (i.e., 
$\rho_2(X)$ is regular aperiodic) if and only if $X$ is definable by 
a first-order formula of $\mathcal{L}_{2,n}$.    
\end{The}

\begin{proof} Let us first show that the condition is sufficient. Let 
$X\subseteq\mathbb{N}$ be a set defined by a formula $\psi$ of 
$\mathcal{L}_{2,n}$. This formula is of the form
$$\psi\equiv (\exists b)(\epsilon_2(b,b)\wedge \varphi)$$
and we can assume that $\psi$ has $n$ as only free variable. 
(If $\psi$ is a sentence, then $X$ is equal to $\mathbb{N}$ or $\emptyset$ 
and the result is obvious.) 
Let us now proceed to some syntaxical transformations. In $\psi$, we 
keep only $\varphi$ in which we replace each occurrence of 
$\epsilon_2(n,x)$, $(\forall x)_2^<$ and $(\exists x)_2^<$ 
with respectively $P_1(x')$, $(\forall x')$ and $(\exists x')$ 
. The remaining variables $x$ are naturally replaced with $x'$. 
It is clear that the obtained formula $\varphi'$ has no 
free variable and belongs to $\mathcal{L}_{SF}$. Indeed, $n$ appears 
in $\varphi$ only through terms of the form $\epsilon_2(n,x)$. 
(As an example, the reader can consider the formulas \eqref{ex2} and 
\eqref{ex1}.)  
Therefore, $\varphi'$ 
defines a star-free language $L$ over $\{0,1\}$. To conclude this part 
of the proof, we have to show that $\rho_2(X)=L$. Let $n$ be such 
that $\langle \mathbb{N},+,\epsilon_2\rangle \models \psi(n)$. 
Assume that $\epsilon_2(n,x)$ appears in 
$\varphi$ with $x=2^l$ for some $l<\log_2 b$ because $x$ is within the scope of a 
quantifier $(\forall x)_2^<$ or $(\exists x)_2^<$. 
It means that $\rho_2(n)$ has a $1$ in the ($l+1$)th position (counting 
positions from 
right to left and beginning with $1$). In $\varphi'$ 
corresponding to $\epsilon_2(n,x)$, we have $P_1(x')$ which means that 
the model of a word --- 
i.e., the representation of an integer --- satisfying $\varphi'$ has a $1$ in 
position $x'$. Thus, we obtain the result when $x'$ is identified as $1+\log_2 x$. 
The upper limit in $\psi$ given by $b$ and appearing in the quantifiers $(\forall 
x)_2^<$ and $(\exists x)_2^<$ is clearly understood in 
$\varphi'$ since in $\mathcal{L}_{SF}$ we consider words as {\it 
finite} models. It 
is the reason for removing the first part of $\psi$ to obtain $\varphi'$ 
and the constant $max$ can be identified as $\log_2 b$. 

Let us now assume that $X$ is a $2$-star-free set. By McNaughton and 
Papert's theorem, $\rho_2(X)$ is defined by a sentence $\varphi$ in 
$\mathcal{L}_{SF}$ where the bound variables are denoted 
$x,y,z,\ldots$ ($\neq n,b$). In $\varphi$, we replace $(\forall x)$, $(\exists 
x)$  and $P_1(x)$ with respectively $(\forall x)_2^<$, $(\exists x)_2^<$ 
and $\epsilon_2(n,x)$ to obtain a formula $\psi'$. It is clear that
$$\psi \equiv (\exists b)(\epsilon_2(b,b) \wedge \psi')$$
is a formula of $\mathcal{L}_{2,n}$ and has possibly $n$ as single 
free variable. To conclude the proof, it is clear that
$$X=\{n\in\mathbb{N}\mid \langle \mathbb{N},+,\epsilon_2\rangle 
\models \psi(n) \}.$$
One can view $b$ as $2^{max}$ if $max$ is a constant of the finite 
model associated to a word. 
\end{proof}

\begin{Exa}{\rm The set $10^*$ can be defined by the following 
formula of $\mathcal{L}_{SF}$
$$
(\exists x) (P_1(x) \wedge (\forall y) (P_1(y)\to x=y)).
$$
The reader can check that this formula corresponds exactly to the 
formula \eqref{ex3} in $\mathcal{L}_{2,n}$ if one proceeds to the 
transformations indicated in the second part of the proof. 
}\end{Exa}

\begin{Rem}{\rm From the logical characterization of the $2$-star-free 
sets given in the 
previous theorem, other equivalent models can be obtained.
In \cite{BHMV}, it is shown how a finite automaton 
$\mathcal{M}$ can 
be effectively derived from a formula $\varphi$ of $\langle 
\mathbb{N},+,V_2\rangle$ defining a $2$-recognizable set 
$X$. Using classical results \cite{Co2}, it is also clear that the characteristic sequence 
of this $X$ is $2$-automatic and the morphisms generating it can be derived 
from $\mathcal{M}$ and thus from $\varphi$.
}\end{Rem}

\section{Generalization to linear numeration systems}
For the sake of simplicity, we have up to now considered the binary 
numeration system but Theorem \ref{thm1} can be extended to more general 
numeration systems.

\begin{Def}{\rm A {\it linear numeration system} $U$  
is a strictly increasing sequence $(U_n)_{n\in\mathbb{N}}$ of 
integers such that $U_0=1$, $\sup \frac{U_{n+1}}{U_n}$ is bounded 
and satisfying for all $n\in\mathbb{N}$ a 
linear recurrence relation 
$$U_{n+k}=c_{k-1} U_{n+k-1}+\cdots +c_0 U_n,\ c_i\in\mathbb{Z},\ c_0\neq 0.$$
By analogy to the binary system, 
the normalized representation of $x$ is denoted by $\rho_U(x)$ (with 
leading zeroes allowed) and 
$V_U(x)$ is the greatest $U_n$ appearing in the greedy decomposition 
of $x$ with a non-zero coefficient ($V_U(0)=U_0=1$). A set $X\subseteq\mathbb{N}$ is 
{\it $U$-recognizable} if $\rho_U(X)$ is regular. 
}\end{Def}

In the following, we shall only consider the class $\mathcal{U}$ 
of linear numeration systems $(U_n)_{n\in\mathbb{N}}$  whose  
 characteristic polynomial is the minimal 
polynomial of a Pisot number. 
For instance, the $k$-ary system and the Fibonacci system belong to 
$\mathcal{U}$. The choice of the class $\mathcal{U}$ relies mainly upon 
the following result. If $U=(U_n)_{n\in\mathbb{N}}$ is a numeration system 
belonging to $\mathcal{U}$ then the 
$U$-recognizable sets are exactly those definable in $\langle \mathbb{N},+,V_U 
\rangle$ (see \cite[Theorem 16]{BH}). In fact, $\mathcal{U}$ is up to 
now the 
largest set of 
numeration systems having well-known and useful properties such as the 
recognizability of addition.

Let $U=(U_n)_{n\in\mathbb{N}}\in\mathcal{U}$. 
Since $\sup \frac{U_{n+1}}{U_n}$ is bounded, the alphabet of the 
normalized representations is finite and is denoted $A_U=\{0,\ldots 
,c\}$. Naturally words over $A_U$ will be interpreted as finite 
models $(M,<,P_1,\ldots,P_c,max)$ and the star-free languages are 
exactly those defined by first-order sentences in this formalism (the 
extension of $\mathcal{L}_{SF}$ defined in Section \ref{sec:SF} is left 
to the reader). As an example, if $w=1230112$ then $P_1=\{2,3,7\}$, 
$P_2=\{1,6\}$ and $P_3=\{5\}$. 

Instead of $V_U(x)$, we shall use $c$ binary relations 
$\epsilon_{j,U}(x,y)$, $j=1,\ldots,c$, meaning that $y$ is an element 
of the sequence $(U_n)_{n\in\mathbb{N}}$ appearing in 
the normalized decomposition of $x$ with a coefficient $j$. Thus
$$x=\sum_{j=1}^c \sum_{\epsilon_{j,U}(x,y)} j\, y.$$

\begin{Rem}{\rm The structures $\langle \mathbb{N},+,V_U 
\rangle$ and $\langle \mathbb{N},+,\epsilon_{1,U},\ldots, 
\epsilon_{c,U} \rangle$  are equivalent, $\epsilon_{j,U}(x,y)$ is  
defined by 
$$(V_U(y)=y)\wedge (\exists t)(\exists z)(x=t+j.y+z \wedge z<y 
\wedge (y<V_U(t)\vee t=0))$$
and $V_U(x)=y$ by
$$(\epsilon_{1,U}(x,y)\vee \cdots \vee \epsilon_{c,U}(x,y)) \wedge 
(\forall z)((\epsilon_{1,U}(x,z)\vee \cdots \vee 
\epsilon_{c,U}(x,z))\to y\le z).$$    
}\end{Rem}
By analogy to $\mathcal{L}_{2,n}$ introduced in the frame of the binary 
system, we can define a language $\mathcal{L}_{U,n}$ of formulas in $\langle 
\mathbb{N},+,\epsilon_{1,U},\ldots, \epsilon_{c,U} \rangle$ having 
possibly a single free variable $n$. For instance, 
$$(\exists x)_U^< \varphi \equiv (\exists x)(\epsilon_{1,U}(x,x)\wedge x<b 
\wedge \varphi).$$
The reader could easily make up the complete definition of $\mathcal{L}_{U,n}$. 

Let us just introduce two notations, if $\varphi$ is any formula of 
$\mathcal{L}_{U,n}$, we shall denote by $\mathfrak{P}(\varphi)$ the 
{\it main part} of 
the formula (i.e., the largest sub-formula in which $b$ is still free), namely the formula is necessarily of the form 
$$\varphi\equiv (\exists b)(\epsilon_{1,U}(b,b)\wedge \mathfrak{P}(\varphi)).$$ 
If $\rho_U(\mathbb{N})$ is aperiodic then it is definable by 
a sentence $\mathfrak{X}$ of $\mathcal{L}_{SF}$. In $\mathfrak{X}$, we replace 
$P_j(x)$, $(\forall x)$ and $(\exists x)$ with $\epsilon_{j,U}(n,x)$, 
$(\forall x)_U^<$ and $(\exists x)_U^<$ respectively to obtain a 
formula $\mathfrak{X}_{\mathbb{N}}$ being the main part of a formula in $\mathcal{L}_{U,n}$.

\begin{The}\label{thm2} Let $U$ be a numeration system in 
$\mathcal{U}$. If $\mathcal{N}_U=0^* \rho_U(\mathbb{N})$ is 
aperiodic then a set $X\subseteq\mathbb{N}$ is $U$-star-free (i.e.,  
$\rho_U(X)$ is regular aperiodic) if and only if $X$ is definable by 
a first-order formula of $\mathcal{L}_{U,n}$ of the form
$$(\exists b)(\epsilon_{1,U}(b,b)\wedge \mathfrak{P}(\varphi) \wedge \mathfrak{X}_{\mathbb{N}})$$
where $\varphi$ is a first-order formula of $\mathcal{L}_{U,n}$.
\end{The}

\begin{proof} The only differences with the proof of Theorem 
\ref{thm1} appear when we show that the condition is sufficient.     
    Roughly speaking, we should have to be careful for the choice of 
a formula $\varphi$ 
in $\mathcal{L}_{U,n}$ because we want to obtain a corresponding formula in 
$\mathcal{L}_{SF}$ valid only for normalized representations 
interpreted as finite models. To avoid this problem we use the 
formula $\mathfrak{X}_{\mathbb{N}}$.    

Let $\varphi$ be any formula of $\mathcal{L}_{U,n}$. It is necessarily of the form 
$$(\exists b)(\epsilon_{1,U}(b,b)\wedge \mathfrak{P}(\varphi)).$$
Assuming that $\mathfrak{X}_{\mathbb{N}}$ and $\mathfrak{P}(\varphi)$ have different 
variables except for $n$ and $b$ then 
$$\psi\equiv (\exists b)(\epsilon_{1,U}(b,b)\wedge \mathfrak{P}(\varphi) \wedge \mathfrak{X}_{\mathbb{N}})$$
is again a formula of $\mathcal{L}_{U,n}$. Adding the part 
$\mathfrak{X}_{\mathbb{N}}$ in such a formula $\psi$ 
ensures that if we transform $\psi$ into a sentence $\psi'$ of 
$\mathcal{L}_{SF}$ (following the scheme given in the proof of 
Theorem \ref{thm1}) then the words satisfying $\psi'$ are all normalized 
$U$-representations and the corresponding language is aperiodic. 
\end{proof}

\begin{Rem}{\rm Notice that for the $k$-ary system, the set of all 
the normalized $k$-representations (allowing leading zeroes) is aperiodic 
 $$0^* \rho_k(\mathbb{N})=\{0,\ldots,k-1\}^*=\overline{\emptyset}$$
 and any word of $\Sigma_k$ is a valid normalized $k$-representation. So in this special 
 case, we do not need 
 a formula $\mathfrak{X}_{\mathbb{N}}$. To be precise, $\mathfrak{X}_{\mathbb{N}}$ is 
 a tautology. In particular, this explains the 
 simpler form of Theorem \ref{thm1} which holds for any numeration 
 system with an integer base $k$.
}\end{Rem}

\begin{Exa}{\rm Let us consider the Fibonacci system given by 
$U_0=1$, $U_1=2$  
and $U_{n+2}=U_{n+1}+U_n$. As a consequence of the greedy algorithm,
$$\mathcal{N}_U= \overline{\overline{\emptyset} 11 
\overline{\emptyset}}$$
is aperiodic and defined by the following sentence of $\mathcal{L}_{SF}$
$$\mathfrak{X}\equiv (\forall x)(\forall y)[(\exists z)(x<z<y) \vee \neg (P_1(x)\wedge 
P_1(y))]$$    
corresponding to
$$\mathfrak{X}_{\mathbb{N}} \equiv (\forall x)_U^<(\forall y)_U^<[(\exists 
z)_U^<(x<z<y) \vee \neg (\epsilon_{1,U}(x)\wedge 
\epsilon_{1,U}(y))].$$ 
So any formula $\varphi$ of $\mathcal{L}_{U,n}$ gives a new formula
$$(\exists b)(\epsilon_{1,U}(b,b)\wedge \mathfrak{P}(\varphi) \wedge \mathfrak{X}_{\mathbb{N}})$$
defining a $U$-star-free subset of $\mathbb{N}$ (which could be finite 
or empty depending on the compatibility of 
the conditions given by $\mathfrak{P}(\varphi)$ and $\mathfrak{X}_{\mathbb{N}}$).

Continuing this example, we show that the set of even integers is not 
$U$-star-free although it is easily definable in 
the Presburger arithmetic by
$$\varphi(n)\equiv (\exists x)(n=x+x).$$
Indeed, $U_n$ is even if and only if $n\equiv 1 \pmod{3}$ and therefore, 
for any $n$, two but not the three words $1(01)^n$,  $1(01)^{n+1}$ and 
$1(01)^{n+2}$ are in the language $\rho_{U}(\mathbb{N})$.  So the set 
of even integers is not definable in $\mathcal{L}_{U,n}$.
}\end{Exa}

\section{Base dependence}\label{sec5}
In this section, we consider once again integer base numeration 
systems and study the base dependence of the star-free property. We 
show that the sets of integers are classified into four categories. 

The proof of the first result in this section does not use the previous 
logical characterization of the $p$-star-free sets but relies mainly on automata theory 
arguments.

\begin{Pro} Let $p,k\ge 2$. A set $X\subseteq \mathbb{N}$ is $p$-star-free if and only 
if it is $p^k$-star-free.
\end{Pro}

\begin{proof}
Let us first show that if $X\subseteq\mathbb{N}$ is $p^k$-star-free then $X$ is $p$-star-free. Assume that  
$\rho_{p^k}(X)$ is 
obtained by an extended regular expression over the 
alphabet $\Sigma_{p^k}=\{0,\ldots,p^k-1\}$ without star operation. 
In this expression, one can replace 
each occurrence of a letter $j \in \Sigma_{p^k}$ with the word 
$0^{k-l}\rho_{p}(j)$ ($l=\vert \rho_{p}(j) \vert$) of 
length $k$. Since we only use 
concatenation product, the resulting expression defining the language 
$L\subset \{0,\ldots,p-1\}^*$ is still star-free and it is clear that $\rho_p(X)=L$.

\begin{Exa}{\rm 
The set $X=\{3.4^n\mid n\in\mathbb{N}\}$ is $4$-star-free, 
$$\rho_{4}(X)=3\, 0^*=\{3\}\, \overline{
\overline{\emptyset}\{1,2,3\} \overline{\emptyset}}.$$ The set $X$ is 
also $2$-star-free, we simply have to replace, $0$, $1$, $2$ and $3$ 
with respectively $00$, $01$, $10$ and $11$ and 
$$\rho_{2}(X)=11\, (00)^*= \{11\}\, \overline{ 
\overline{\emptyset}\{01,10,11\} \overline{\emptyset}}.$$    
}\end{Exa}

Before continuing the proof, we recall another characterization of 
the star-free languages given by McNaughton and Papert.
\begin{Def}{\rm A deterministic finite automaton is {\it permutation 
free} if there is no word that makes a nontrivial permutation (i.e., 
not the identity permutation) of any subset of the set of states.  In 
the same way, a language is said to be {\it permutation free} if its minimal 
automaton is permutation 
    free.
}\end{Def}

\begin{Exa}{\rm  The automaton depicted in Figure \ref{fig:1} is not 
permutation free.
\begin{figure}[h!tbp]
    \begin{center}
	\includegraphics{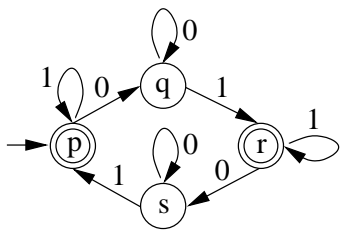}
	\caption{A non permutation free automaton.}\label{fig:1}
	\end{center}
\end{figure}
Indeed, the word $01$ makes a non trivial permutation of the set 
$\{p,r\}$.
}\end{Exa}

\begin{The}\cite[Theorem 5.1]{NP}\label{char} A language is star-free 
if and only if it is permutation free.
\end{The}

Let us also recall a well-known result from automata theory.
\begin{Pro}\cite[Section III.5]{Ei}\label{minimal} Let $L \subset 
\Sigma^*$ be a 
regular language having $\mathcal{M}_L=(Q_L,q_0,F_L,\Sigma,\delta_L)$ as 
minimal automaton. If $\mathcal{M}=(Q,q_0',F,\Sigma,\delta)$ is an 
accessible deterministic automaton recognizing $L$ then there exists 
an application $\Phi:Q\to Q_L$ such that 
$\Phi$ is onto and  for each $q\in Q$ and $w\in \Sigma^*$, 
$$\Phi(\delta(q,w))=\delta_L(\Phi(q),w).$$
\end{Pro}

Assume now that $X\subseteq\mathbb{N}$ is $p$-star-free. Using Remark 
\ref{zero} and Theorem \ref{char}, $0^*\rho_p(X)$ is a permutation 
free language and we denote by $\mathcal{M}=(Q,q_0,F,\Sigma_p,\delta)$ 
its minimal automaton. From  $\mathcal{M}$, we build a new automaton 
$\mathcal{M}'=(Q,q_0,F,\Sigma_{p^k},\delta')$ having the same set of 
states. The transition 
function $\delta'$ of $\mathcal{M}'$ is defined as follows. For each  
$j\in\Sigma_{p^k}$, $p,q \in Q$, let $w=0^{k-l} \rho_p(j)$ where 
$l=\vert \rho_p(j) \vert$, then $\delta'(p,j)=q$ if and only if 
$\delta(p,w)=q$.

\begin{Exa}{\rm Let $\Sigma_2=\{0,1\}$ and consider the automaton 
$\mathcal{M}$ depicted in Figure \ref{fig:2}.
\begin{figure}[h!tbp]
    \begin{center}
	\includegraphics{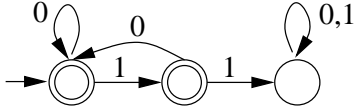}
	\caption{An automaton $\mathcal{M}$ over $\Sigma_2$.}\label{fig:2}
	\end{center}
\end{figure}
If we consider the $4$-ary numeration system, we build a new 
automaton $\mathcal{M}'$ depicted in Figure \ref{fig:3} by considering in 
$\mathcal{M}$ the paths of label $00$, $01$, $10$ and $11$ replaced 
respectively by $0$, $1$, $2$ and $3$.
\begin{figure}[h!tbp]
    \begin{center}
	\includegraphics{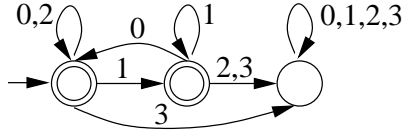}
	\caption{The corresponding automaton $\mathcal{M}'$ over $\Sigma_4$.}\label{fig:3}
	\end{center}
\end{figure}
}\end{Exa}

The automaton $\mathcal{M}'$ is accessible. Let $q\in Q$. Since 
$\mathcal{M}$ is accessible, there exists a word $w\in\Sigma_p^*$ such 
that $\delta(q_0,w)=q$. Observe that in $\mathcal{M}$, we have a loop 
of label $0$ in the initial state $q_0$. So for any $n\in\mathbb{N}$, 
$\delta(q_0,0^n w)=q$. Choose $n$ such that $\vert 0^n w \vert=i k$. 
The word $0^n w$ corresponds to a word $w'\in (\Sigma_{p^k})^i$ and 
$\delta'(q_0,w')=q$. So $\mathcal{M}'$ is accessible.

The automaton $\mathcal{M}'$ is permutation free. Assume the contrary. 
If there exists $T\subseteq Q$ and a word $w'\in\Sigma_{p^k}^*$ such 
that $w'$ makes a nontrivial permutation of $T$. Then $w'$ corresponds 
to a word $w\in\Sigma_p^*$ producing a nontrivial permutation in the 
same subset $T$ of the set of states of $\mathcal{M}$. This is a 
contradiction.

It is clear from the construction of $\mathcal{M}'$, that this 
automaton accepts exactly the language $0^*\rho_{p^k}(X)$. The only 
remaining problem before being able to apply Theorem \ref{char} 
is that $\mathcal{M}'$ is not necessarily reduced. Indeed, due to its 
definition, through $\mathcal{M}'$ we only look in $\mathcal{M}$ at 
words of length $ik$, $i\ge 0$.

Let $\mathcal{M}''=(Q'',q_0'',F'',\Sigma_{p^k},\delta'')$ 
be the minimal automaton of $0^*\rho_{p^k}(X)$. 
Thanks to Proposition \ref{minimal}, we have an onto application 
$\Phi$ between the automata $\mathcal{M}'$ and $\mathcal{M}''$ namely 
between the sets $Q$ and $Q''$. 
To conclude the proof by applying Theorem \ref{char}, 
we have to show that $\mathcal{M}''$ is permutation free.  
Assume that there exists a subset $T$ of $Q''$ 
and a word $w\in\Sigma_{p^k}^*$ such that 
$w$ makes a nontrivial permutation of $T$. So there exists a state 
$q\in T$ such that $\delta''(q,w)\in T$ and $\delta''(q,w)\neq q$. 
Therefore $\mathcal{M}'$ is not permutation free, $w$ makes a 
nontrivial permutation of $\Phi^{-1}(T) \subseteq Q$. Indeed, there 
exist $r\in \Phi^{-1}(q) \subseteq \Phi^{-1}(T)$ and $s \in 
\Phi^{-1}(\delta''(q,w)) \subseteq \Phi^{-1}(T)$ such that $r\neq s$ 
and $\delta'(r,w)=s$. This is a contradiction because we have shown 
previously that $\mathcal{M}'$ is permutation free.
\end{proof}

It is well-known that any finite union of arithmetic progressions is $p$-star-free for 
some $p$ (see \cite[Theorem 1.4]{dLR}). So a natural question 
is to determine if an arithmetic progression $r+ s\, 
\mathbb{N}=\{r+sn\mid n\in \mathbb{N}\}$, $s>1$, 
is $p$-star-free for any $p\ge 2$ or only for some specific bases $p$. 

\begin{Exa}{\rm The set of even integers is $2$-star-free and therefore 
$2^n$-star-free for each $n$. But this set is not $3$-star-free, 
indeed $\rho_3(2\mathbb{N})$ is the set of words over $\{0,1,2\}$ 
having an even number of $1$ (and therefore the minimal automaton of 
this language is not counter-free, which is another way to say that 
the language is not permutation free). 
Notice also that $2$ and $10$ are 
multiplicatively independent but $2\mathbb{N}$ is $10$-star-free. 
Actually, it is easy to see that the set of even integers is 
$(2p)$-star-free, for any $p$. So with this example, we see that we 
obtain a slighty different phenomenom that the one encountered in 
Cobham's theorem.
}\end{Exa}

A finite union of arithmetic progressions being ultimately periodic, 
we can always write it as $\cup_{j=1}^q (r_j+ s\, \mathbb{N}) \cup F$ where $F$ is a 
finite set and $s$ is the l.c.m. of the periods of the different progressions. 
Since aperiodicity is preserved up to a finite modification of a 
language, we can forget the finite set $F$ and assume that $r_j<s$. 
Union of aperiodic sets being again aperiodic, we shall consider a 
single set $r+ s\, \mathbb{N}$.

\begin{Pro} The set $r+ s\, \mathbb{N}$, (with 
$r<s$ and $s>1$) is $(is)$-star-free for any integer $i>0$.
\end{Pro}

\begin{proof} The reader can easily check that the language made up 
of the $(is)$-ary expansions of the elements in $r+ s\, \mathbb{N}$ is  
    $$\Sigma_{is}^* \{r,r+s,\ldots ,r+(i-1)s\}$$
which is a definite\footnote{A language $L\subset\Sigma^*$ is said to 
be {\it definite} if 
there exist finite languages $M$ and $N$ such that $L=N\cup \Sigma^* M$. 
So to test the membership of a word in $L$, we only have to look at its 
last letters.} language.
\end{proof}

We even have a better situation.
\begin{Pro}\label{pro:pa} Let $r+ s\, \mathbb{N}$ be such that $r<s$, $s>1$ and the 
factorization of $s$ as a product of primes is of the form
$$s=p_1^{\alpha_1}\cdots p_k^{\alpha_k},\ \alpha_i>0$$
If $P=\Pi_{j=1}^k p_j$ then $r+ s\, \mathbb{N}$ is $(iP)$-star-free 
for any integer $i>0$.
\end{Pro}

\begin{proof} Let $\alpha=\sup_{j=1,\ldots,k} \alpha_j$. By definition 
of $P$, it is clear that $(iP)^{\alpha+n}$ is a multiple of $s$ for 
all integers $n\ge 0$ and $i>0$. So in the 
$(iP)$-ary expansion of an integer the digits corresponding to those 
powers of $iP$ provide the decomposition with multiples of $s$. To 
obtain an element of $r+ s\, \mathbb{N}$, we thus have to focus on 
the last $\alpha$ digits corresponding to the powers $1$, $iP$, \ldots 
,$(iP)^{\alpha-1}$ of weakest weight. Consider the finite set  
$$Y=\{r+ns\mid n\in\mathbb{N}\ {\rm and }\ r+ns<(iP)^\alpha\}.$$
For each $y_j\in Y$, $j=1,\ldots ,t$, consider the word $\rho_{iP}(y_j)$ preceded by some 
zeroes to obtain a word $y_j'\in\Sigma_{iP}^*$ of length $\alpha$. To 
conclude the proof, observe that the language made up 
of the $(iP)$-ary expansions of the elements in $r+ s\, \mathbb{N}$ is
$$\Sigma_{iP}^* \, \{y_1',\ldots ,y_t'\}$$
and is a definite language.
\end{proof}

\begin{Rem}{\rm The situation of Proposition \ref{pro:pa} cannot be 
improved. Indeed with the previous notations, consider an integer $Q$ 
which is a product of some but not all the prime factors appearing in $s$. 
For the sake of simplicity, assume that 
$$Q=p_2^{\beta_2}\cdots p_k^{\beta_k},\ 1\le \beta_j \le \alpha_j.$$
For each $n$, $Q^n \not\equiv 0 \pmod{s}$. Indeed, if $Q^n=i\, s$ then 
$$p_2^{n \beta_2}\cdots p_k^{n \beta_k}=i \, p_1^{\alpha_1}\cdots 
p_k^{\alpha_k}$$ which is a contradiction since $p_1$ does not appear 
in the left hand side factorization. Moreover, it is clear that the 
sequence $(Q_n \mod s)_{n\in\mathbb{N}}$ is ultimately periodic. 
Therefore $\rho_Q(r+ s\, \mathbb{N})$ is regular but not star-free 
because, due to this 
periodicity, the corresponding automaton is not counter-free. As an 
example, one can check that $6\mathbb{N}$ is neither $2$-star-free 
nor $3$-star-free.
}\end{Rem}
    
To summarize the situation, the sets of integers can be classified 
into four categories:
\begin{enumerate}
\item The finite and cofinite sets are $p$-star-free for any 
 $p>1$. 
\item The ultimately periodic sets of period $s=p_1^{\alpha_1}\cdots 
p_k^{\alpha_k} >1$ are $(iP)$-star-free for $P=\Pi_{j=1}^k p_j$ and any 
$i>0$. In particular, these sets are $P^m$-star-free for $m\ge 1$.
\item Thanks to Cobham's theorem, if a $p$-recognizable set $X$ is not a 
finite union of arithmetic progressions then  $X$ is only 
$p^k$-recognizable for $k\ge 1$ ($p$ being simple\footnote{Being 
multiplicatively dependent is an equivalence relation over 
$\mathbb{N}$, the smallest 
element in an equivalence class is said to be {\it simple}. For instance, 
$2,3,5,6,7,10,11$ are simple.}). So if a 
$p$-star-free set $X$ is not ultimately periodic then  $X$ is only 
$p^k$-star-free for $k\ge 1$ ($p$ being simple).
\item Finally, there are sets which are not $p$-star-free for any $p>1$.
\end{enumerate}

\section{p-adic number systems}
The $p$-ary numeration system is built upon the the sequence 
$U_n=p^n$ and the representation of an integer is a word over the 
alphabet $\{0,\ldots,p-1\}$ computed through the greedy algorithm. 
On the other hand, the $p$-adic numeration system is built upon the same sequence but 
representations are written over the alphabet $\{1,\ldots,p\}$. It can 
be shown that each integer has a unique $p$-adic representation (see 
\cite{S1} for an exposition on $p$-adic number systems). 
For instance, the use of $p$-adic system may be relevant to 
remove the ambiguity due to the presence of leading zeroes in a 
$p$-ary representation. Indeed, $0$ is not a valid digit in a $p$-adic 
representation (see for instance \cite[p. 303]{KRS} for a relation 
to L systems).

In this 
small section, we show that the $p$-star-free sets are 
exactly the sets of integers whose $p$-adic representations made up a 
star-free language. 

Capital Greek letters will represent finite alphabets.

\begin{figure}[h!tbp]
    \begin{center}
	\includegraphics{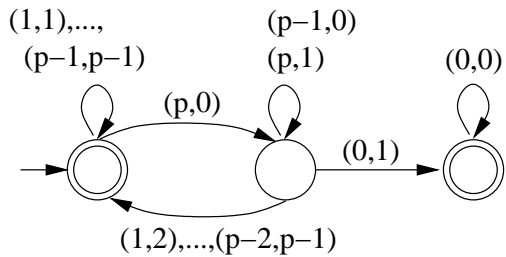}
	\caption{From $p$-adic to $p$-ary representation.}\label{fig:4}
	\end{center}
\end{figure}

If $\Delta\subset\mathbb{Z}$ is a finite alphabet and $w=w_n\cdots w_0$ 
is a finite word over $\Delta$, we denote by $\pi_p(w)$ the {\it 
numerical value} of $w$,
$$\pi_p(w)=\sum_{i=0}^n w_i\, p^i.$$
For instance, $1001$ and $121$ are respectively the $2$-ary and 
$2$-adic representations of $9$,
$$\pi_2(1001)= \pi_2(121)=9.$$
Let $w\in\Delta^*$ be such that $\pi_p(w)\in\mathbb{N}$. The 
partial function $\nu_p:\Delta^*\to \{0,\ldots,p-1\}^*$ 
mapping $w$ onto $\rho_p(\pi_p(w))$ is called the {\it normalization function}. 
Thanks to a result of C.~Frougny, the graph of this function is 
regular whatever the alphabet $\Delta$ is \cite{Fr}. For the case we are 
interested in, the language
$$\widehat{\nu_p}^R=\{(u,v)\mid u\in\{1,\ldots,p\}^*0^*, v\in 
\Sigma_p^*, \vert u\vert=\vert v\vert , \pi_p(u^R)=\pi_p(v^R) \}$$
is the reversal of the graph of the normalization funciton mapping the $p$-adic 
representation of an integer onto its $p$-ary representation. The 
trim minimal automaton (the sink has not been represented) of 
$\widehat{\nu_p}^R$ is given in Figure \ref{fig:4} and is clearly permutation 
free. 
So thanks to Theorem \ref{char}, $\widehat{\nu_p}^R$ is star-free.

\begin{Lem}\label{ll:1} A language $L\subseteq \Sigma^*$ is star-free if and only 
if $L^R$ is star-free. 
\end{Lem}

\begin{proof} Let $u,v,w\in \Sigma^*$. Assume that $L$ is aperiodic, 
for $n$ large enough
\begin{eqnarray*}
    u\, v^nw\in L^R & \Leftrightarrow & w^R (v^R)^n u^R \in L \\
     & \Leftrightarrow & w^R (v^R)^{n+1} u^R \in L \\
     & \Leftrightarrow & u\, v^{n+1} w \in L^R. 
\end{eqnarray*}
\end{proof}
Thus $\widehat{\nu_p}$ is a star-free language. 

We denote by $p_1$ and $p_2$ the canonical homomorphisms of 
projection, if $A$ and $B$ are sets, $p_1:A\times B \to A : (a,b)\mapsto a$ and $p_2:A\times 
B \to B : (a,b)\mapsto b$.

\begin{Lem}\label{ll:2} Let $L\subseteq \Sigma^*$ be a star-free language. Then 
the language
$$L\oplus \Gamma^*=\{(x,y)\mid x\in L, y\in \Gamma^*, \vert x\vert 
=\vert y\vert \}$$ is also star-free.
\end{Lem}

\begin{proof} Let $u,v,w\in (\Sigma\times \Gamma)^*$. Since $L$ is 
star-free and $p_1$ and $p_2$ are letter-to-letter (length preserving 
homomorphisms), we have
    \begin{eqnarray*}
    u\, v^nw\in L\oplus \Gamma^* & \Leftrightarrow & p_1(u\, v^nw) \in 
    L \ \& \  p_2(u\, v^nw) \in \Gamma^* \\
    & \Leftrightarrow & p_1(u)\, p_1(v)^n p_1(w) \in 
    L \ \& \  p_2(u)\, p_2(v)^n p_2(w) \in \Gamma^* \\
    & \Leftrightarrow & p_1(u)\, p_1(v)^{n+1} p_1(w) \in 
    L \ \& \  p_2(u)\, p_2(v)^{n+1} p_2(w) \in \Gamma^* \\
    & \Leftrightarrow & p_1(u\, v^{n+1}w) \in 
    L \ \& \  p_2(u\, v^{n+1}w) \in \Gamma^* \\
    & \Leftrightarrow & u\, v^{n+1}w\in L\oplus \Gamma^*\\
\end{eqnarray*}
\end{proof}   
Naturally, we can also define the language $\Gamma^* \oplus L$ in a 
similar way.

Generally, the homomorphic image of a star-free language is not 
star-free \cite[p. 12]{NP} but the following weaker result holds.

\begin{Lem}\label{ll:3} If a language $L\subset (\Sigma\times \Gamma)^*$ of 
couples of words of the same length is star-free then 
$p_1(L)\subset\Sigma^*$ and $p_2(L)\subset\Gamma^*$ are also 
star-free.
\end{Lem}

\begin{proof} One can apply the same reasoning as the one given in the proof of 
 the previous lemma.
\end{proof}

We are now able to prove the main result of this section (notice that 
a small mention to $p$-adic systems appears also in \cite{dLR2}).
\begin{Pro} A set $X\subseteq \mathbb{N} $ is $p$-star-free if and 
only if the language made up of the $p$-adic representations of the 
elements in $X$ is star-free.
\end{Pro}

\begin{proof} If $\rho_p(X)$ is a star-free language then thanks to 
Lemma \ref{ll:2}, 
$$\{0,\ldots ,p\}^* \oplus \rho_p(X) $$
is star-free. Thanks to Lemma \ref{ll:1} and since the family of 
aperiodic languages is closed under boolean operations, the language
$$L=[\{0,\ldots ,p\}^* \oplus \rho_p(X) ]\cap \widehat{\nu_p}$$
is again star-free. To conclude the first part of the proof, we apply 
Lemma \ref{ll:3}, $p_1(L)$ is star-free and it is clear that this 
language is exactly made up of the $p$-adic representations of the 
elements in $X$. 

Conversely, let $M\subset \{1,\ldots,p\}^*$ be such that 
$\pi_p(M)=X$. If $M$ is star-free then  thanks to Remark \ref{zero} 
and using the previous lemmas, 
$$p_2[(0^*M\oplus \{0,\ldots,p-1\}^*)\cap \widehat{\nu_p}]$$
is star-free.
\end{proof}

\section{Acknowledgments}
The author warmly thanks Antonio Restivo which has suggested this work 
during the thematic term {\it Semigroups, Algorithms, Automata and 
Languages} in Coimbra.



\begin{thebibliography}{99}
\bibitem{BH} V. Bruy\`{e}re, G. Hansel, Bertrand numeration systems and recognizability. 
Latin American Theoretical INformatics (Valpara\'{\i}so, 1995). 
{\it Theoret. Comput. Sci.} {\bf 181} (1997), 17--43.
\bibitem{BHMV} V. Bruy\`{e}re, G. Hansel, C. Michaux, R. Villemaire, 
 Logic and $p$-recognizable sets of integers. Journ\'{e}es
 Montoises (Mons, 1992). {\it Bull. Belg. Math. Soc. Simon Stevin} 
 {\bf 1} (1994), 191--238.
\bibitem{Cob} A. Cobham, On the base-dependence of sets of 
numbers recognizable by finite automata. {\it Math. Systems Theory} 
{\bf 3} (1969) 186--192.
\bibitem{Co2} A. Cobham, Uniform tag sequences. 
{\it Math. Systems Theory} {\bf 6} (1972), 164--192. 
\bibitem{Ei} S. Eilenberg, {\it Automata, languages, and machines}. Vol. A. 
Pure and Applied Mathematics, Vol. 58. Academic Press, New York, (1974).
\bibitem{F} A. S. Fraenkel, Systems of numeration. {\it Amer. Math. Monthly} 
{\bf 92} (1985), 105--114.
\bibitem{Fr} C. Frougny, Representations of numbers and finite automata. 
{\it Math. Systems Theory} {\bf 25} (1992), 37--60.
\bibitem{Hod} B. R. Hodgson, D\'ecidabilit\'e par automate fini. 
 {\it Ann. Sci. Math. Qu{\'e}bec} {\bf 7} (1983), 39--57.
\bibitem{KRS} L. Kari, G. Rozenberg, A. Salomaa, L systems. in {\it  
Handbook of formal languages}, Vol. 1, 253--328, Springer, Berlin, (1997).
\bibitem{dLR} A. de Luca, A. Restivo, Star-free sets of integers. 
{\it Theoret. Comput. Sci.} {\bf 43} (1986), 265--275.
\bibitem{dLR2} A. de Luca, A. Restivo, Representations of integers 
and laguage theory, Mathematical foundations of computer science (Prague, 1984), 407--415, 
{\it Lecture Notes in Comput. Sci.} {\bf 176}, Springer, Berlin, 
(1984). 
\bibitem{NP} R. McNaughton, S. Papert, {\it Counter-free automata}. 
 M.I.T. Research Monograph, No. 65. The M.I.T. Press, Cambridge, 
 Mass.-London, (1971).
\bibitem{S1} A. Salomaa, {\it Formal languages}. Academic Press, New 
York, (1973).
\bibitem{S} M. P. Sch\"{u}tzenberger,  
 On finite monoids having only trivial subgroups. 
{\it Information and Control} {\bf 8} (1965), 190-194. 
\bibitem{STT} H. Straubing, D. ThŽrien, W. Thomas,
 Regular languages defined with generalized quantifiers. 
{\it Inform. and Comput.} {\bf 118} (1995), 289--301.
\bibitem{T} W. Thomas, Classifying regular events in symbolic logic. 
 {\it J. Comput. System Sci.} {\bf 25} (1982), 360--376.
\bibitem{V}  R. Villemaire, The theory of $\langle \mathbb{N},+,V_k,V_l\rangle$ 
is undecidable. {\it Theoret. Comput. Sci.} {\bf 106} (1992), 337--349.

\end{thebibliography}
\end{document}